\documentclass[preprint,showpacs,nofootinbib]{revtex4}
\usepackage{mathrsfs}
\usepackage{float,epsfig}
\usepackage{dcolumn}
\usepackage{bm}
\usepackage{graphicx}
\usepackage{bm}
\usepackage{subfigure}
\usepackage{amsmath,amssymb,amsthm}
\usepackage[colorlinks=true,linkcolor=blue]{hyperref}

\begin{document}
\title{\bf Phase transition and Thermodynamical geometry of Reissner-Nordstr\"om-AdS Black Holes in Extended Phase Space}

\author{Jia-Lin Zhang$^{1,2}$,  Rong-Gen Cai$^{1,2}$ and Hongwei Yu$^{1,3}$}
\affiliation{$^1$ Department of Physics and Key Laboratory of Low Dimensional Quantum Structures and Quantum Control of Ministry of Education, Hunan Normal University, Changsha 410081, China\\
$^2$ State Key Laboratory of Theoretical Physics, Institute of Theoretical Physics,
Chinese Academy of Sciences, Beijing 100190, China \\
$^3$ Center for Nonlinear Science and Department of Physics, Ningbo University, Ningbo 315211, China}

\begin{abstract}

We study the thermodynamics and thermodynamic geometry of a five-dimensional Reissner-Nordstr\"om-AdS black hole in the extended phase space
by treating the cosmological constant as being related to the number of colors in the
boundary gauge theory and its conjugate quantity as the associated
chemical potential. It is  found that the contribution of the charge
of the black hole to the chemical potential is always positive and the
existence of charge make the chemical potential become positive more
easily. We calculate the scalar curvatures of the thermodynamical
Weinhold metric, Ruppeiner metric and Quevedo metric, respectively,
in the fixed $N^2$ case and the fixed $q$ case. It is found
that in the fixed $N^2$ case the divergence of the scalar curvature
is related to the divergence of the specific heat with fixed electric
potential in the Weinhold metric and Ruppeiner metric, and the
divergence of the scalar curvature  in the Quevedo metric  corresponds to
the divergence of the specific heat with fixed electric charge density.
 In the fixed $q$ case,  however, the divergence of the scalar
curvature is related to the divergence of the specific heat with fixed
chemical potential in the Weinhold metric and Ruppeiner metric,
while in the Quevedo metric the divergence of the scalar curvature corresponds
 to the divergence of the specific heat with fixed number of
colors and the vanishing of the specific heat with fixed chemical
potential.
\end{abstract}
\pacs{04.60.-m, 04.70.Dy, 11.25.-w} \maketitle

\section{Introduction}

The  well-known AdS/CFT correspondence relates a
weakly coupled gravitational theory in $d$-dimensional anti-de Sitter (AdS) spacetime
to a strongly coupled conformal field theory (CFT) in a
$(d-1)$-dimensional boundary of the AdS
space~\cite{Gubser:1998bc,Maldacena:1997re,Witten:1998qj,Witten:1998qj2}
(for a review, see~\cite{Aharony}).   In the spirit of the AdS/CFT correspondence, the Hawking-Page phase transition  between the stable large black hole
and thermal gas in AdS space~\cite{Hawking} can be  interpreted as the confinement/deconfinement
phase transition in the dual strongly coupled gauge theory~\cite{Witten:1998qj2}.  Indeed in the past years we have witnessed increasing interest in studying
thermodynamics and the phase structure of black holes in AdS space.

 For a Kerr-Newmann black hole in general relativity,  Davies~\cite{Davies} found that
some heat capacities  diverge at some values of black hole
parameters, and he argued that some second order phase transitions
will happen in the Kerr-Newmann black hole. For a
Reissner-Nordstr\"om AdS (RN-AdS) black hole, such a phase
transition was studied in some details in~\cite{Myers}. In a
canonical ensemble with a fixed charge, in particular,  it was found
that there exists a phase transition between small and large black
holes. This phase transition behaves very like the gas/liquid phase
transition in a Van der Waals system~\cite{Myers,rgc1}. However, an
complete identification between a RN-AdS black hole and the Van der
Waals system was recently realized in~\cite{Kubiznak}, where the
negative cosmological constant plays the role as pressure, while its
conjugate acts as the thermodynamic volume of the black hole in the
so-called extended phase space~\cite{Dolan1,Dolan2} (For a recent
review, see \cite{Dolan3}). Recently,  there has been a lot of works
studying the thermodynamics and phase transition in the extended
phase space for black holes in AdS space (For an  incomplete  list
of reference see~\cite{others}).

In the  AdS/CFT correspondence, the negative cosmological constant
is related to the degrees of freedom of the dual CFT. Thus  an interesting
question arises as to whether the interpretation of the cosmological constant as pressure is
applicable to the boundary CFT. Very recently, it was argued that it is more suitable to view
the cosmological constant as the number of colors in the dual gauge field and its conjugate as associated chemical
potential~\cite{Dolan2014,Johnson,Kastor:2014dra}. This interpretation was examined for ${\cal N}$=4 supersymmetric
 Yang-Mills theory at large $N$ in \cite{Dolan2014,ZCY}, by studying the corresponding thermodynamics of a Schwarzschild-AdS black hole. They calculated the chemical potential conjugate to the number of colors, and found that the chemical potential in the high temperature phase of the Yang-Mills theory is negative and decreases as temperature increases.
  For spherical black holes in the bulk the chemical potential approaches zero as the temperature is lowered below
  the Hawking-Page temperature and changes its sign at a temperature near the temperature at which the heat capacity diverges. This phenomenon might be related to the
  Bose-Einstein condensation in the dual field theory~\cite{Dolan2014}. Furthermore,  in \cite{Kastor:2014dra} the authors  studied the associated chemical potential  from the point of view
  of holographic entanglement entropy.

  On the other hand,  recently a lot of attention has been attracted to applying the thermodynamical geometry to thermodynamics and
  phase transition of black holes.  The geometrical ideas to ordinary
thermodynamical systems  were first introduced by  Weinhold~\cite{Weinhold}. He considered
 a kind of metric defined as the second derivatives
of internal energy with respect to entropy and other extensive
quantities for a thermodynamic system. Based on the
fluctuation theory of equilibrium thermodynamics,
Ruppeiner~\cite{Ruppeiner-2} introduced another metric
defined as the minus second derivatives of
entropy  with respect to the internal energy and other extensive
quantities.  He argued that the scalar curvature of the Ruppeiner metric
can reveal the micro interaction of the system and its divergence  is
 related to a certain phase transition~\cite{Rupreview}. In fact  the
Weinhold metric is conformal to the Ruppeiner
metric with the inverse temperature as the conformal factor~\cite{Salamon}. Unfortunately, both the Weinhold metric and
Ruppeiner metric are  not invariant under Legendre transformation. 
A few years ago Quevedo {\it et al.} \cite{Quevedo-1,Quevedo-2,Quevedo-3,Quevedo-4}
proposed an approach to obtain a new metric which is Legendre invariant in the
space of equilibrium state. As far as we know, applying
the thermodynamical geometry to black hole thermodynamics  started in \cite{Ferrara:1997tw}, which shows that the
Weinhold metric is proportional to the metric on the moduli space for supersymmetric extremal
black holes with vanishing Hawking temperature, and the Ruppeiner metric governing fluctuations
naively diverges, which is consistent with the fact that near the extremal limit, the thermodynamical description of black holes should be invalid.
Applying the thermodynamical geometry approach to the phase transition of black holes was followed in \cite{CC,Aman:2003ug}, and for more recent references see the review paper~\cite{Rup2013}
and references therein.
In particular,  Ref.~\cite{Mansoori:2013pna} has investigated  the relation between the divergence of the scalar curvature of thermodynamical geometry in different ensembles
 and the singularity of heat capacities.  In a previous paper we have studied thermodynamical geometry and the phase transition for a Schwarzschild AdS black hole in the extended phase space where the cosmological constant
is related to the number of colors of dual gauge field~\cite{ZCY}.

In this paper, we will extend the previous study to the case of a
charged black hole in AdS space, and concretely we will study the
thermodynamics and thermodynamical geometry for a five-dimensional
RN-AdS black hole by viewing the number of colors as a
thermodynamical variable from the viewpoint of dual CFT.  For this,
we will calculate energy density and entropy density for the dual
CFT and then obtain the chemical potential associated with the
number of colors in the next section. In Sec.~\ref{III}
and~\ref{IV}, we will calculate the thermodynamical curvatures of
the Weinhold metric, Ruppeiner metric and Quevedo metric,
respectively, for the thermodynamical system in the fixed $N^2$ case
and the fixed $q$ case, in order to see the relation between the
thermodynamical curvature and the phase transition of black holes in
AdS space.  We end the paper with conclusions in Sec.~\ref{V}.

\section{Thermodynamics of RN-AdS black holes in extended phase space}
\label{II}
Let us start with the following  Einstein-Maxwell theory with a negative cosmological constant  in five-dimensional
 spacetime
 \begin{equation}
 S=\frac{1}{16\pi{G_5}}\int_M{d^5}x\sqrt{-g}\bigg[R-L^2F^2+\frac{12}{L^2}\bigg],\;
 \end{equation}
 where $L$ is the AdS radius related to the cosmological constant
as $\Lambda=-6/L^2$.  As shown in \cite{Myers}, the above action with an additional  Chern-Simons term can be viewed as an effective truncation of type IIB supergravity on an $S^5$.
The action admits  a five-dimensional  RN-AdS black  hole as its exact solution, which can be
written  in the static coordinates as
\begin{equation}
\label{RN}
ds_5^2=-f(r)dt^2+\frac{1}{f(r)}dr^2+r^2h_{ij}dx^idx^j,
\end{equation}
where $h_{ij} dx^idx^j$ is the line element of a three-dimensional Einstein space $\Sigma_3$ with constant curvature $6k$, and the metric function $f(r)$ is given by~\cite{Myers}
\begin{equation}\label{fr}
f(r)=k-\frac{m}{{r^2}}+\frac{r^2}{L^2}+\frac{q_L^2}{r^4}\;.
\end{equation}
The integration constant $m$ is related to the mass of the black
hole
 \begin{equation}\label{adsmass}
 M=\frac{ 3\omega_3 }{16 \pi G_5}m
 \end{equation}
where $\omega_3$ denotes the volume of $\Sigma_3$, while the parameter
$q_L$ has a relation to the physical charge  $Q$ of the black hole as
\begin{equation}
q_L=\frac{4\pi  G_5 Q L}{\sqrt{3}\omega_3}\;.
\end{equation}
Without loss of generality, one can take the scalar curvature
parameter $k$ of the three-dimensional space $\Sigma_3$ as $k=1$,
$0$, or $-1$, respectively.  Uplifting this solution (\ref{RN}) to ten dimensions, one has~\cite{Myers}
\begin{equation}
\label{ten}
ds_{10}^2 = ds_5^2 + L^2  \sum^3_{i=1} [d\mu_i^2 +\mu_i^2 (d\varphi_i+\frac{2}{\sqrt{3}}A_{\mu}dx^{\mu})^2],
\end{equation}
where $\mu=0,\cdots,4$, the variables $\mu_i$ are direction cosines
on $S^5$ satisfying $\sum^3_{i=1}\mu_i^2=1$, and the $\varphi_i$ are
rotation angles on $S^5$. The ten-dimensional spacetime (\ref{ten})
can be viewed as the near horizon geometry of $N$ rotating black
$D3$-branes in type IIB supergravity. In that case, the AdS radius
$L$ has a relation to the number $N$ of
D3-branes~\cite{Maldacena:1997re}
\begin{equation}\label{n1}
L^4=\frac{\sqrt{2}N\ell_p^4}{\pi^2} \equiv \alpha^2 N,
\end{equation}
where $\ell_p$ is the ten-dimensional Planck length. According to  the AdS/CFT correspondence, the spacetime (\ref{ten}) can be
regarded as the gravity dual to ${\cal N}$=4 supersymmetric Yang-Mills theory in the Coulomb branch. Then  $N$ is nothing, but the rank of the gauge group of the supersymmetric $SU(N)$ Yang-Mills Theory. In the large $N$ limit, the number of degrees of freedom of the ${\cal N}$=4 supersymmetric Yang-Mills theory is proportional to $N^2$~\cite{Gubser}.

 The  black hole  horizon
$r_h$  is determined by  equation $f(r)=0$ by taking the largest
real root of the equation. Then with Eq.~(\ref{fr}) and
Eq.~(\ref{adsmass}), the mass of black hole can be expressed as
\begin{equation}\label{m1}
M=\frac{ 3\omega_3 }{16 \pi G_5}\big(k
r_h^2+\frac{r_h^4}{L^2}+\frac{4G_5^2Q^2L^2}{3\pi^2r_h^2}\big)\;.
\end{equation}
 Using the Bekenstein-Hawking entropy formula of black hole, we
have the black hole entropy
\begin{equation}
S=\frac{A}{4G_5}=\frac{\omega_3 r_h^3}{4G_5}\;.
\end{equation}
Note that $G_5=G_{10}/(\pi^3L^5)$ and $G_{10}=\ell_p^8$.
Furthermore, let us notice that the dual CFT to the RN-AdS black
hole lives in the AdS boundary with a metric (up to a conformal
factor)
\begin{equation}
ds^2 =- dt^2 +L^2 h_{ij}dx^idx^j.
\end{equation}
Namely the CFT lives in a space with volume $V_3=\omega_3 L^3$. We
see that the volume depends on the number of colors $N^2$. In order
to remove the effect of volume change when one varies the number of
colors, let us consider the  corresponding densities of some
thermodynamic quantities of the dual CFT as
\begin{eqnarray}\label{rho}
&& \rho = \frac{M}{V_3}=\frac{3 \pi^2 L^2}{16 G_{10}}r_h^2 (k+ \frac{r_h^2}{L^2})+\frac{G_{10}Q^2}{4L^6\pi^6r _h^2}, \\
&& s= \frac{S}{V_3}=\frac{\pi^3}{4G_{10}}L^2 r_h^3,\\
&& q=\frac{Q}{V_3}=\frac{Q}{2\pi^2L^3}\;.
\end{eqnarray}
The Hawking temperature of the black hole can be given by requiring the absence of the potential conical singularity of the
Euclidean black hole at the horizon. A simple calculation gives
\begin{equation}
\label{temper} T = \frac{1}{2\pi r_h}\left ( k
+2\frac{r_h^2}{L^2}\right)-\frac{2 G_{10}^2 Q^2}{3 L^8 \pi^9r_h^5}.
\end{equation}
The energy density can be expressed in terms of the entropy density, the
number of colors  and the  charge density as
\begin{equation}\label{rho2}
\rho= \frac{3 D\pi^2}{16 G_{10}}
 \left [ \alpha k N^{1/6}s^{2/3} +N^{-2/3}s^{4/3}D\right]+\frac{D_1}{2}N^{1/3}
 q^2s^{-2/3}\;,
 \end{equation}
 where we have introduced  $D \equiv \big[{4G_{10}}/(\pi^3\alpha)\big]^{2/3}$ and $D_1 \equiv \big(G_{10}\alpha^2/2\big)^{1/3}$ for convenience in the following discussions.
According to the standard thermodynamic relations, the corresponding
intensive variables of the CFT can be calculated. For example,
the temperature of the CFT is
\begin{equation}\label{temperature}
T=\left (\frac{\partial \rho}{\partial s}\right)_{N^2,q}=
  \frac{D\pi^2}{8 G_{10}}
  \left(\alpha k N^{1/6} s^{-1/3} + 2 N^{-2/3}
  s^{1/3}D\right)-\frac{D_1}{3}N^{1/3}q^2s^{-5/3},
  \end{equation}
which is noting, but just the Hawking temperature Eq.~(\ref{temper}) of the
black hole. In order to keep that the RN-AdS black hole describes a thermal state of the dual field theory,  the Hawking temperature must be non-negative,
i.e., the following condition has to be satisfied
\begin{equation}\label{T-uneq}
3D\pi ^2s^{4/3}\left(2 D s^{2/3}+k N^{5/6} \alpha
\right)\geq8D_1G_{10}N q^2\;.
\end{equation}
Here the equality means an
extremal black hole with vanishing temperature $T=0$. The  static electric potential associated with the charge of the black hole can also be obtained as
\begin{equation}\label{elec-potential}
\Phi=\left (\frac{\partial \rho}{\partial q}\right)_{s,N^2}= D_1q
N^{1/3} s^{-2/3}\;.
\end{equation}
This can be explained by the chemical potential associated with R current in the dual Supersymmetric  Yang-Mills  theory.  But we remind  the readers not to be  confused with the following chemical potential
associated with the number of colors in the dual field theory.
The chemical potential $\mu$ conjugate to the number of colors
is defined as~\cite{Dolan2014}
\begin{equation}
\label{chem} \mu = \left(\frac{\partial \rho}{\partial
N^2}\right)_{s,q}= \frac{ D\pi^2}{16 G_{10}} \bigg(\frac{1}{4}
\alpha k N^{-11/6} s^{2/3}
-N^{-8/3}s^{4/3}D\bigg)+\frac{D_1}{12}q^2N^{-5/3}s^{-2/3},
\end{equation}
which is the  measure of the energy cost to the system of increasing
the number of colors.  We see that the contribution of
electric charge to the chemical potential  is always positive.

As  a result, we have the first law of thermodynamics
\begin{equation}
\label{firstlaw}
d\rho = Tds +\mu dN^2+\Phi{dq}.
\end{equation}
Clearly one can see that the case relating the cosmological constant to the number of colors is
quite different from the case viewing the cosmological constant as the pressure.  In the former case,  one can see from (\ref{firstlaw}) that
the mass of the black hole can be viewed as the  internal energy of the system, while in the latter case, the mass of the black hole is
taken as the  enthalpy~\cite{Dolan1,Dolan2}.

For the cases of $k=0$ and $k=-1$, it is easy to see that from
Eq.~(\ref{temperature}) for fixed $N^2$ and $q$, the Hawking
temperature increases monotonically with the entropy density $s$.
Besides, with the  inequality~(\ref{T-uneq}) for the
case of $k=0$ or $k=-1$, it is not difficult to deduce from
Eq.~(\ref{chem}) that the chemical potential is always negative for
non-extremal black holes. This is consistent with our understanding
that the chemical potential for classical gas is negative and will
become more negative as temperature increases~\cite{Callen}. Therefore, the cases
of $k=0$ and $k=-1$ are trivial for studying the phase transition
and thermodynamic geometry. We will focus our attention on the
case of $k=1$ in what follows.

\begin{figure*}[htbp]
\centering{\subfigure[]{\label{a1}
\includegraphics[scale=0.8]{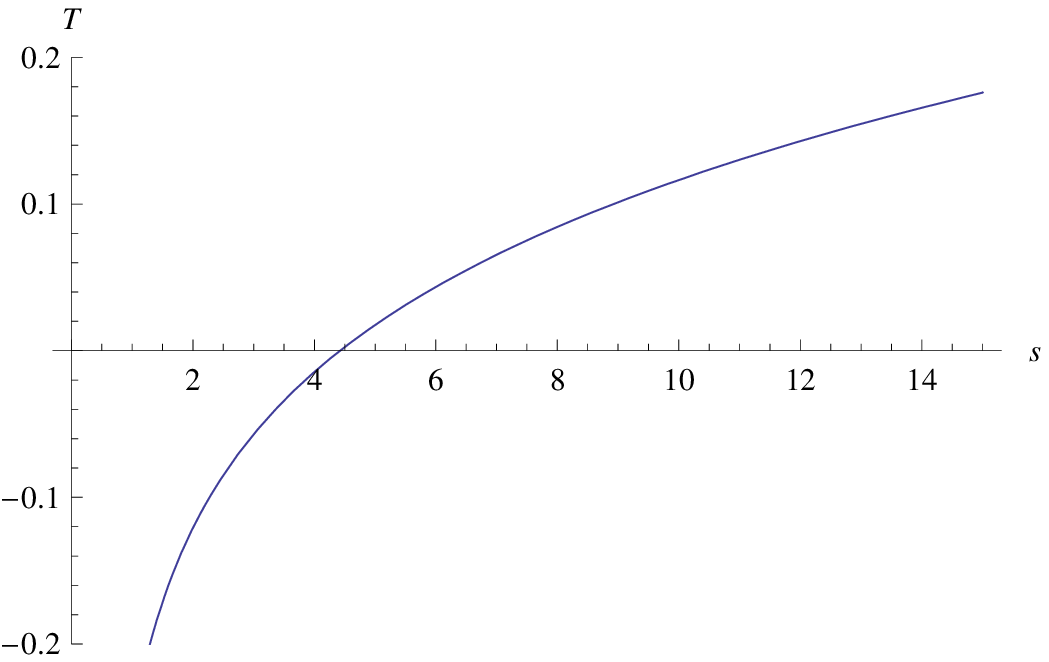}}\subfigure[]{\label{b1}\includegraphics[scale=0.8]{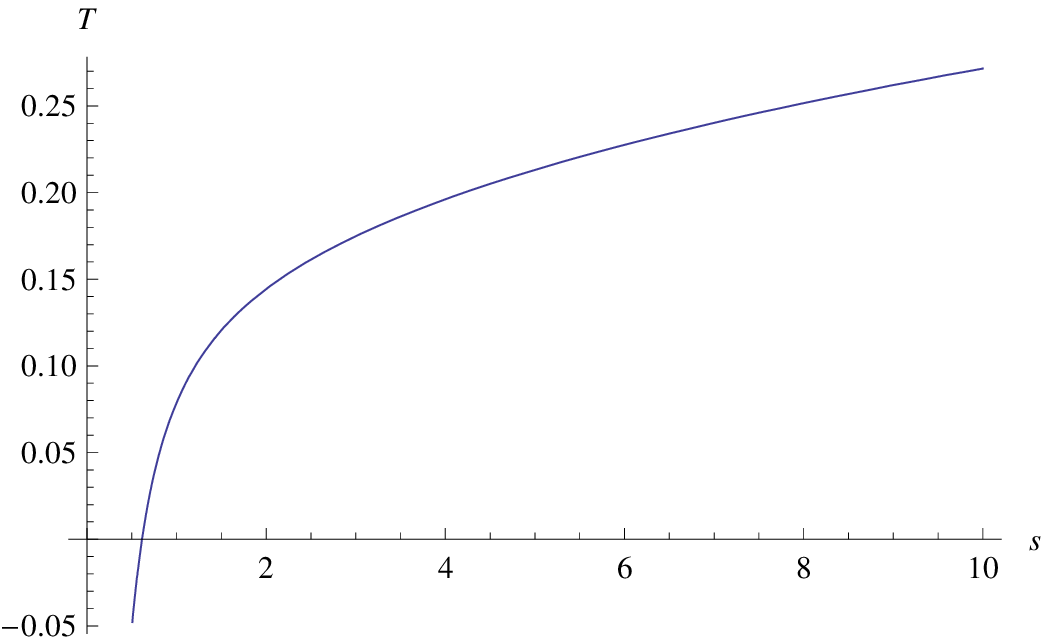}}
\subfigure[]{\label{c1}\includegraphics[scale=0.8]{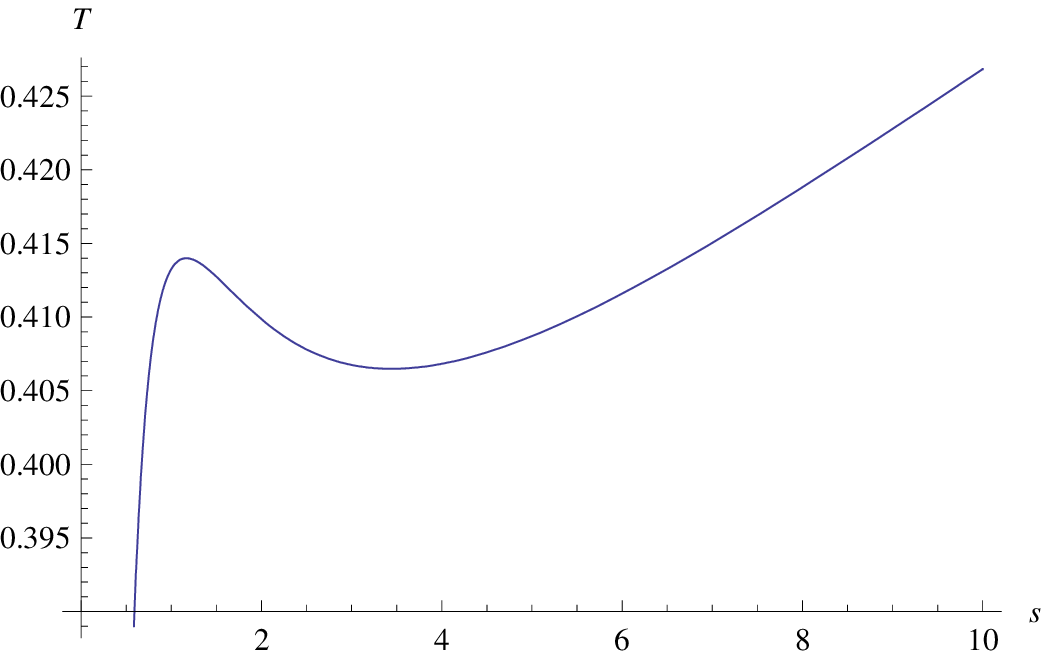}}}
\caption{The Hawking temperature vs  entropy density  $s$ for (a)$k=-1\;$
(b)$k=0\;$ (c)$k=1 \;$ with $N=10,\;$ $q=0.4$,  and
$\ell_p=1$.}\label{temp-k}
\end{figure*}

For the case of $k=1$, we have
\begin{equation}\label{T-partial}
\frac{\partial{T}}{\partial{s}}=\frac{3D\pi
^2s^{4/3}\left(2Ds^{2/3}-N^{5/6}\alpha\right)+40N q^2
D_1G_{10}}{72N^{2/3}s^{8/3}G_{10}}\;.
\end{equation}
Then we can conclude that the Hawking temperature may not be a monotonic
 function of $s$ by choosing other proper thermodynamic
variables. For a fixed number of colors, there  is a  critical point
of $q$, which can be determined by
\begin{equation}
\bigg(\frac{\partial{T}}{\partial{s}}\bigg)_{q_{crit}}=\bigg(\frac{\partial^2{T}}{\partial{s^2}}\bigg)_{q_{crit}}=0\;.
\end{equation}
It is easy to obtain  that the  critical point satisfies
\begin{equation}\label{q-crit}
q^2_{crit}=\frac{N^{3/2}\pi^2 \alpha^3}{360DD_1G_{10}}\;.
\end{equation}
When $q>q_{crit}\;,$  the Hawking temperature is always a monotonically
increasing function of $s\;.$  Similarly, if we fix the charge
density $q$, there must be
 a critical point for the number of colors. For the case of
$N<N_{crit}
=\frac{12}{\alpha^2}\big(\frac{75D^2q^4D_1^2G_{10}^2}{\pi^4}\big)^{1/3}\;,$
the Hawking temperature is always a monotonically increasing
function of $s\;,$ and otherwise, there must exist  a certain interval
of $s$ in which the Hawking temperature  is  a monotonically
decreasing function. In Fig.~(\ref{temp-k}) we show the behavior of
the temperature in the cases of $k=-1$, $0$ and $1$, respectively.
We can see that in the case of $k=1$, the temperature in the
interval $(1.167,3.443)$ is a monotonically decreasing function of
$s$ with parameters $N=10\;$,  $q=0.4$ and $\ell_p=1\;$.

The Helmholtz  free energy density can be calculated as
\begin{eqnarray}
\label{gibbs} {\cal F }&=& \rho-Ts =\frac{5 N^{1/3}q^2 D_1}{6
s^{2/3}}-\frac{D^2\pi^2 s^{4/3}}{16 N^{2/3} G_{10}}+\frac{D N^{1/6}
\pi ^2 s^{2/3} \alpha }{16 G_{10}}\;.
\end{eqnarray}
Note that the free energy
also has a chance to be positive, if the following condition is satisfied
\begin{equation}
\frac{3D\pi^2s^{4/3}\left(2Ds^{2/3}+N^{5/6}\alpha\right)}{8 N D_1
G_{10}}\geq{q}^2>\frac{3D\pi^2s^{4/3}\left(D s^{2/3}-N^{5/6} \alpha
\right)}{40 ND_1G_{10}}\;.
\end{equation}
The sign change of  the free energy  indicates  the appearance of the Hawking-Page phase transition.
\begin{figure}[htbp]
\centering{\subfigure[ $N=10$]{\label{a1}
\includegraphics[scale=0.7]{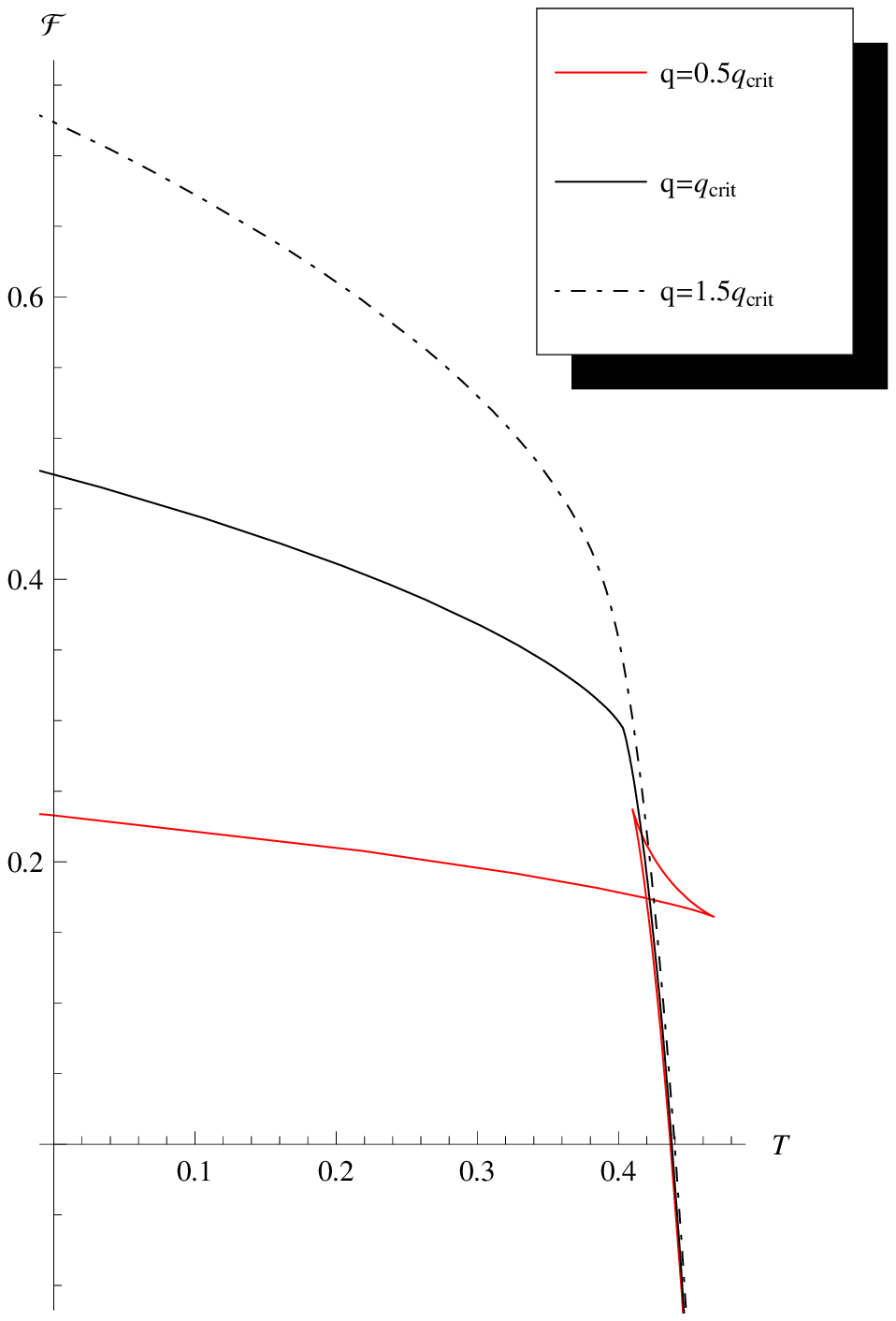}}\subfigure[ $q=0.4$]{\label{b1}
\includegraphics[scale=0.7]{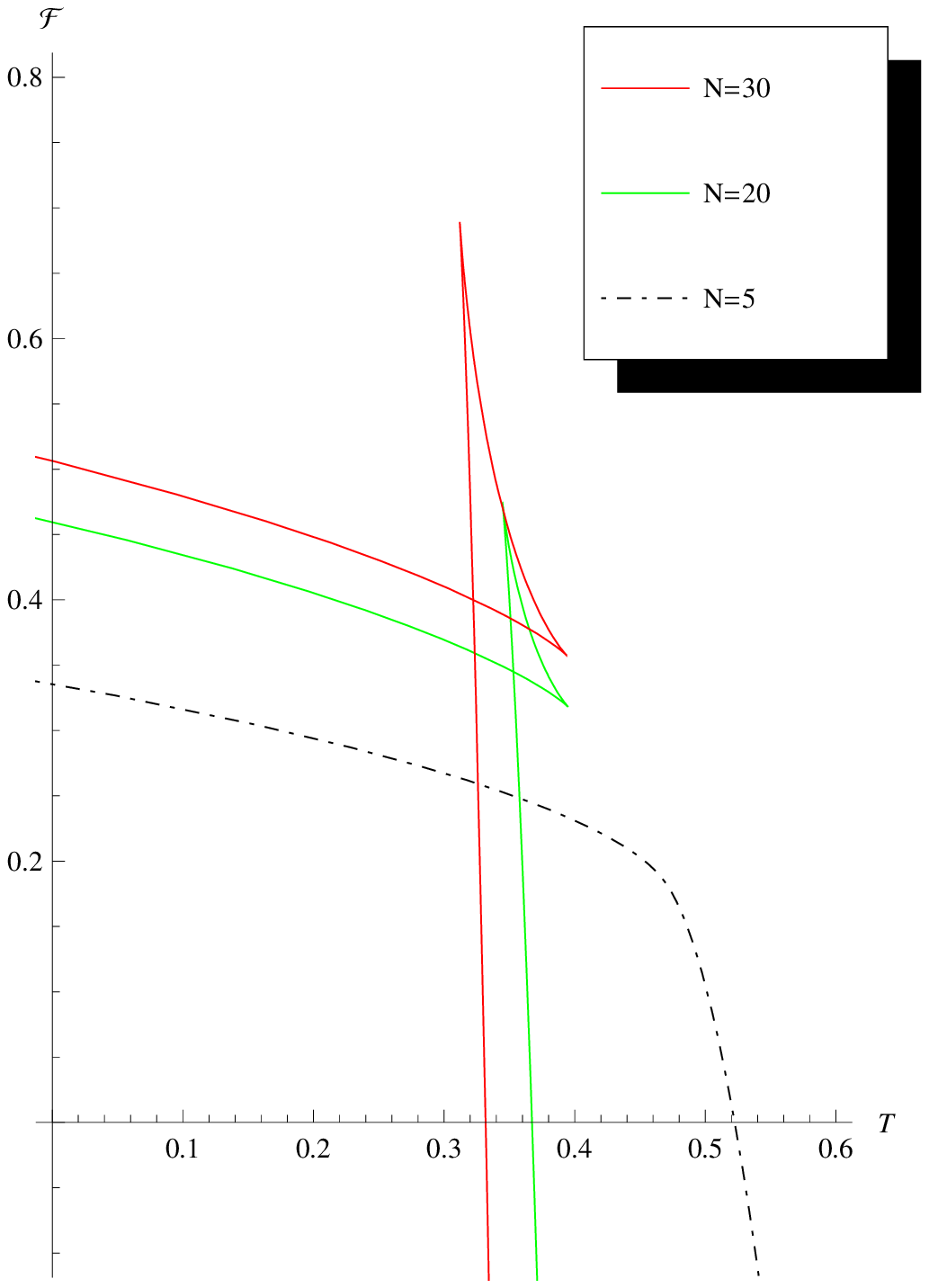}}}
\caption{The Helmholtz  free energy as a function of Hawking
temperature for (a) a fixed number of colors and (b) a fixed charge
density with  $k=1$ and $\ell_p=1$. The ``swallow tail'' appears in
figure (a) for $q<q_{crit}\approx0.4817$ and in figure (b) for
$N>N_{crit}=\frac{12}{\alpha^2}\big(\frac{75D^2q^4D_1^2G_{10}^2}{\pi^4}\big)^{1/3}\approx7.805\;.$
}\label{Gibbs}
\end{figure}
 In Fig.~(\ref{Gibbs}), we plot the free energy with
 respect to the Hawking temperature. It can be seen that the ``swallow tail", a type signal for the first order phase transition,
 appears in both the cases with a fixed number of colors and a fixed charge.  The phase transition between the small black hole and the large one in the former case has been studied in
 \cite{Myers},  while the latter case is the new finding in this paper.

Now we turn to the case of the chemical potential conjugate to the number of colors. According to
Eq.~(\ref{chem}) and inequality~(\ref{T-uneq}), the condition for a
positive chemical potential can be written as
\begin{equation}
\frac{3D\pi^2s^{4/3}\left(2Ds^{2/3}+N^{5/6}\alpha\right)}{8 N D_1
G_{10}}\geq{q}^2>\frac{3D\pi^2s^{4/3}\left(4Ds^{2/3}-N^{5/6}\alpha\right)}{16
N D_1 G_{10}}\;\;.
\end{equation}

\begin{figure*}[htbp]
\centering
\includegraphics[scale=1]{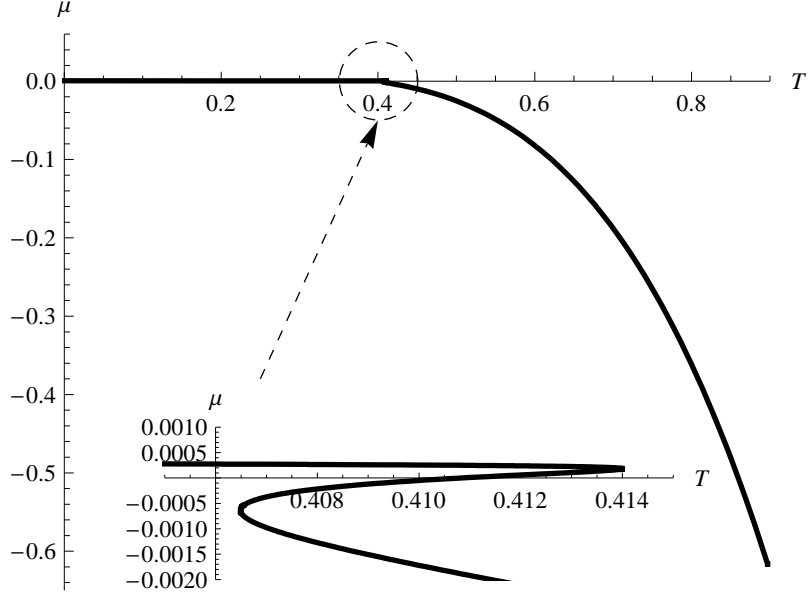}
\caption{The chemical potential vs the Hawking temperature  with
parameters  $k=1\;,N=10\;,q=0.4$ and $\ell_p=1\;.$ }\label{mu-T}
\end{figure*}

\begin{figure}[htbp]
\centering
\includegraphics[scale=0.8]{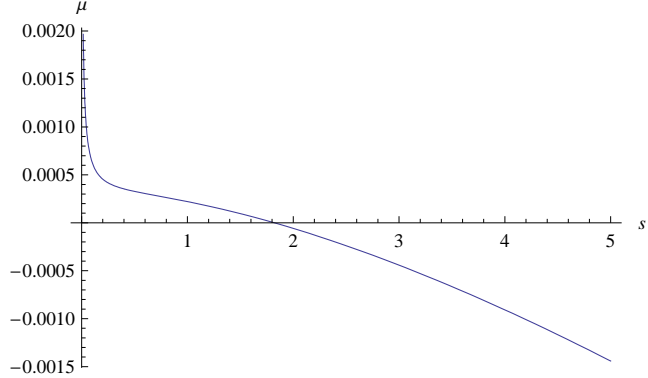}
\caption{The chemical potential as a function of $s$.
Here  we take  $N=10\;,q=0.4\;,k=1\;$ and $\ell_p=1\;.$
}\label{mu-s}
\end{figure}
In Fig.~(\ref{mu-T}), we  show  the chemical potential as a function
of temperature $T$ for fixed $q$ and $N\;,$ while  in
Fig.~(\ref{mu-s}) the chemical potential is plotted  as a function
of $s$ for fixed $N=10$ and $q=0.4\;.$  As we can see that the
chemical potential surely has a chance to be positive by choosing
proper thermodynamic variables. It should be pointed out that when
the chemical potential approaches zero and becomes positive, quantum
effects should play some role~\cite{Dolan2014}. The existence of the
electric charge of black hole makes the chemical potential go beyond
zero more easily on some level.  We can see from Fig.~(\ref{mu-T})
that there exists a multi-valued region, which just corresponds to
the unstable region of the black hole with a negative heat capacity
(see plot (c) in Fig.~(\ref{temp-k})), while Fig.~(\ref{mu-s}) shows
that the chemical potential is positive for small black holes.

 In the
following sections, we will study the phase transition and
thermodynamical geometry of the RN-AdS black hole in the fixed $N^2$
case and  the fixed $q$ case, respectively.

\section{ Phase transition and  thermodynamic
geometry of the RN-AdS black hole in the fixed $N^2$ case } \label{III}
\begin{figure*}[htbp]
\centering{
\includegraphics[scale=1]{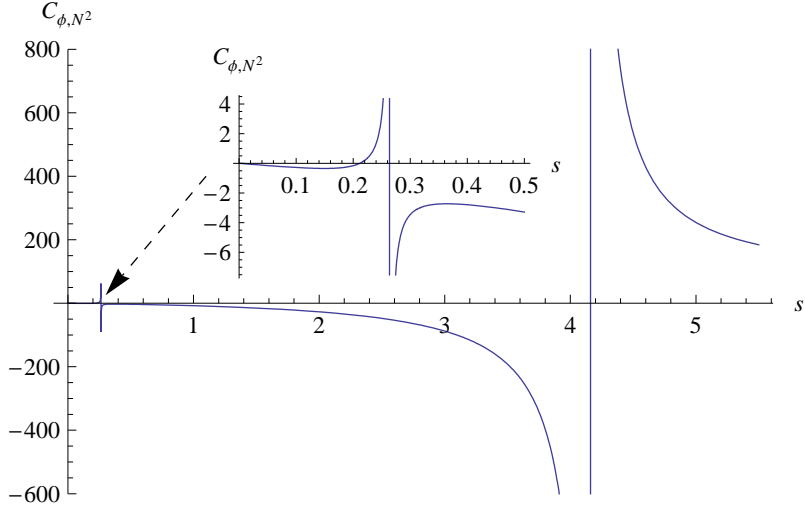}}
\caption{The specific heat $C_{\Phi,N^2}$  of the  RN-AdS black hole
with respect to entropy density $s$ for parameters
$k=1\;,N=10\;,q=0.4$ and $\ell_p=1\;.$ There are two divergent
points at $s_1\approx0.264$ and $s_2\approx4.160$  in this case. It
should be pointed out that $s<0.211$ corresponds the situation of
negative Hawking temperature, this nonphysical situation is not  our
main attention and not considered detailed in following paragraphs
and figures.}\label{cphns}
\end{figure*}

\begin{figure*}[htbp]
\centering{
 \subfigure[$\;q=0.4<q_{crit}$]{\label{a1}
\includegraphics[scale=0.8]{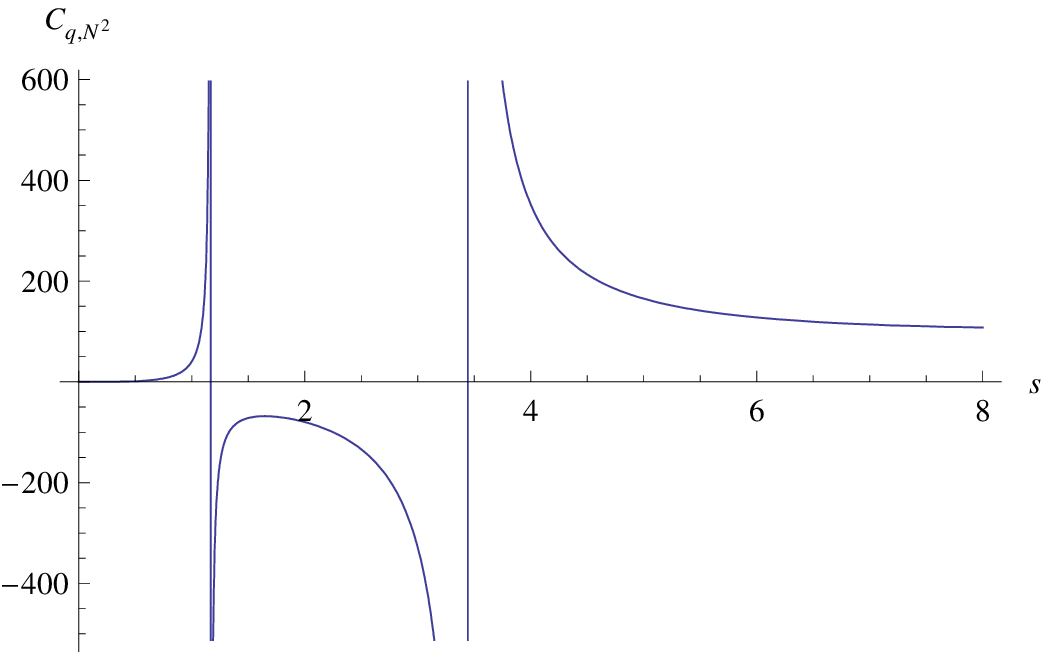}}\subfigure[$\;q=q_{crit}$]{\label{b1}
\includegraphics[scale=0.8]{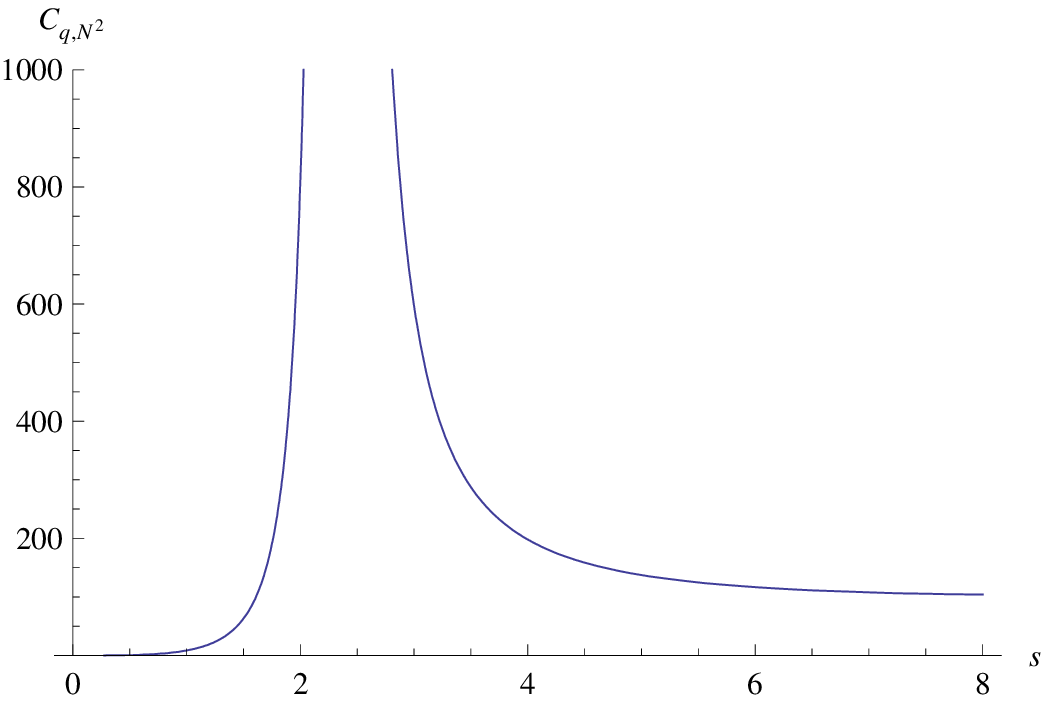}}\\{\subfigure[$\;q=1>q_{crit}$]{\label{c1}
\includegraphics[scale=0.8]{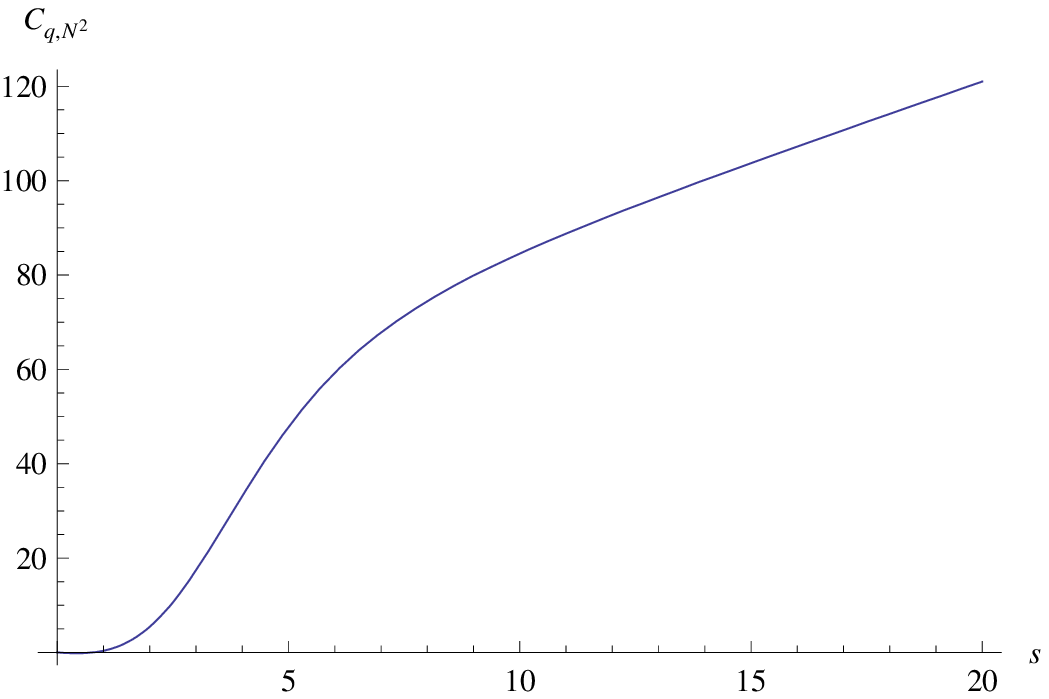}}}}
\caption{The specific heat $C_{q,N^2}$ of the RN-AdS black hole with
respect to entropy density for parameters  $k=1\;,N=10$ and
$\ell_p=1\;.$ (a):  Two divergences at $s_3\approx1.167$ and
$s_4\approx3.443\;,$(b): One divergence at  $s_5\approx2.339\;,$(c)
: No divergence.}\label{cqns}
\end{figure*}

In this section, we will study the phase transition and
thermodynamic  geometry of the black hole in the ensemble with a fixed number of
the colors. Once the number of the colors is fixed, it means that
the cosmological constant is kept at a certain value and not treated as a
thermodynamic variable. The corresponding phase transition and
thermodynamics of RN-AdS black holes have been studied extensively in
Refs.~\cite{Myers,rgc1,YuTian2012}. Here we pay attention to the relation between the phase transition and thermodynamic geometry. In this case, the associated  specific heats
can be calculated as

\begin{equation}\label{c-nq}
C_{q,N^2}=T\bigg(\frac{\partial{s}}{\partial{T}}\bigg)_{q,N^2}=\frac{3
s \left[3 D \pi^2s^{4/3}\left(2Ds^{2/3}+N^{5/6} \alpha \right)-8 N
q^2 D_1 G_{10}\right]}{3 D \pi ^2 s^{4/3} \left(2D s^{2/3}-N^{5/6}
\alpha\right)+40Nq^2D_1 G_{10}}\;,
\end{equation}
and
\begin{equation}\label{c-nphi}
C_{\Phi,N^2}=T\bigg(\frac{\partial{s}}{\partial{T}}\bigg)_{\Phi,N^2}=\frac{3
s \left[3 D\pi^2s^{4/3} \left(2 D s^{2/3}+N^{5/6}\alpha \right)-8 N
q^2D_1G_{10}\right]}{3 D \pi ^2 s^{4/3} \left(2 D s^{2/3}-N^{5/6}
\alpha \right)+8Nq^2D_1G_{10}}\;.
\end{equation}
 In the
canonical ensemble with  a fixed $N^2\;,$ a critical point exists
and it can be calculated by Eq.~(\ref{q-crit}). The behavior of the
$C_{\Phi,N^2}$  as a function of $s$ is plotted in
Fig.~(\ref{cphns}) and the specific heat $C_{q,N^2}$ is  shown in
Fig.~(\ref{cqns}). The behavior of $C_{q,N}$ is consistent with the
one of the temperature shown in Fig.~(\ref{temp-k})(c): for small
and large black holes the specific heat is positive, while it is
negative for the intermediate black holes when $q$ is less than the
critical value, while the specific heat is always positive in the
case when the charge $q$ is larger than the critical one.


Now we turn to the thermodynamical geometry of the black hole  to
see whether the thermodynamical curvature can reveal the singularity
of these two specific heats.  The Weinhold metric~\cite{Weinhold} is
defined as the second derivatives of internal energy with respect to
entropy and other extensive quantities in the energy representation,
while the Ruppeiner metric \cite{Ruppeiner-2} is  related to the
Weinhold metric by a conformal factor of temperature~\cite{Salamon}
\begin{equation}\label{conform}
ds^2_R=\frac{1}{T}ds^2_W\;.
\end{equation}
The Weinhold metric and  Ruppeiner metric, which are
 dependent on the choice of thermodynamic potentials, are not
Legendre invariant, while  the Legendre invariant Quevedo metric is defined as~\cite{Quevedo-1,Quevedo-2,Quevedo-3,Quevedo-4}
\begin{equation}\label{Gq}
g=\bigg(E^c\frac{\partial{\phi}}{\partial{E^c}}\bigg)
\bigg(\eta_{ab}\delta^{bc}\frac{\partial^2{\phi}}{\partial{E^c\partial{E^d}}}dE^adE^d\bigg)\;, \ \eta_{cd}={\rm diag}(-1,1,\ldots,1)
\end{equation}
where $\phi$ denotes the thermodynamic potential, $E^a$ and $I^a$
respectively represent the set of extensive variables and the set of
intensive variables, and $a=1,2, \cdots, n$.

Now we calculate the thermodynamical curvature for the RN-AdS black
hole.  The Weinhold metric is given by
\begin{equation}\label{gW1}
g^W=\Bigg(\begin{matrix}\rho_{ss} & \rho_{sq}\\
{\rho_{qs}}&{\rho}_{qq}\end{matrix}\Bigg)\;,
\end{equation}
where $\rho_{ij}$ stands for  $\partial^2{\rho}/\partial x^i\partial
x^j$, and $x^1=s$, $x^2=q$. The scalar curvature  of this metric can
be calculated directly. Substituting Eq.~(\ref{rho2}) into
Eq.~(\ref{gW1}) leads to the scalar curvature:
\begin{equation}\label{RW}
R^W=-\frac{48 D N^{3/2} \pi ^2 s^2 \alpha G_{10}}{\big[3D\pi^2
s^{4/3}\left (2Ds^{2/3}-N^{5/6}\alpha \right) + 8N
q^2D_1G_{10}\big]^2}\;.
\end{equation}
On the other hand, by considering Eq.~(\ref{conform}), the Ruppeiner metric can be written
as
\begin{equation}\label{gR}
g^R=\frac{1}{T}\Bigg(\begin{matrix}\rho_{ss} &
\rho_{sq}\\{\rho}_{qs}& {\rho_{qq}}\end{matrix}\Bigg)\;,
\end{equation}
and the corresponding curvature of this metric is
\begin{equation}\label{RR}
R^R=\frac{A_1(s,q)}{B_1(s,q)E_1(s,q)}\;,
\end{equation}
where $A_1(s,q)$ is a regular function  without any singular behavior, and
 the other two auxiliary functions are given by
\begin{eqnarray}
B_1(s,q)&=&\big[3D\pi^2s^{4/3}\left(2Ds^{2/3}-N^{5/6} \alpha
\right)+8G_{10}Nq^2D_1\big]^2\nonumber\\
E_1(s,q)&=&3D\pi^2 s^{4/3}\left (2 D s^{2/3} + N^{5/6}\alpha \right)
-8 N q^2 D_ 1 G_ {10}\;.\nonumber
\end{eqnarray}

\begin{figure*}[htbp]
\centering{\subfigure[Weinhold metric]{\label{a1}
\includegraphics[scale=0.8]{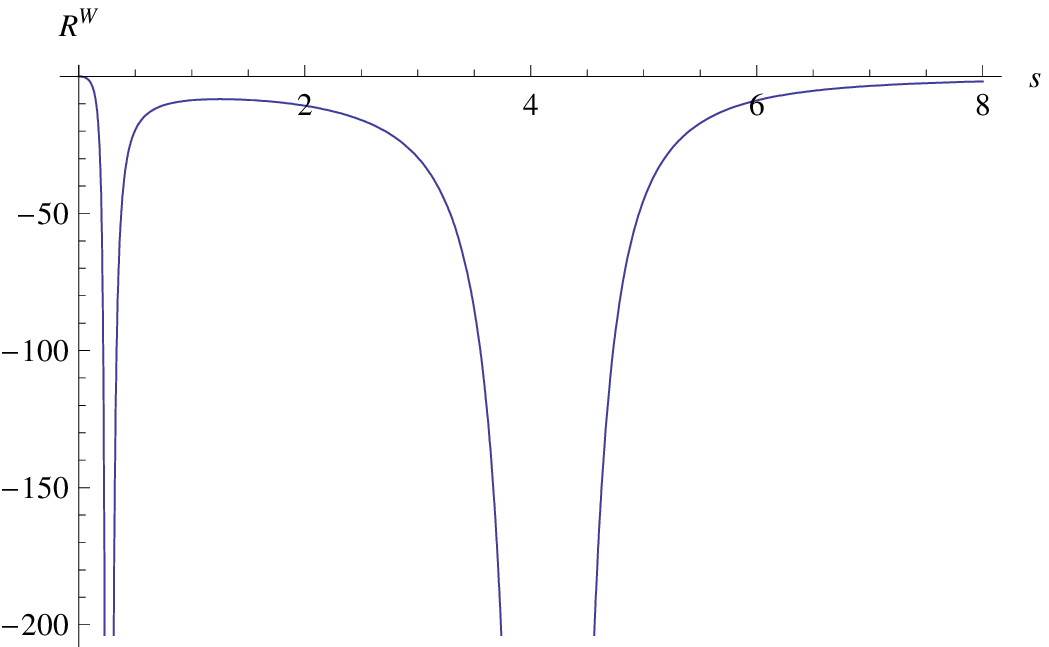}}\subfigure[Ruppeiner metric]{\label{b1}\includegraphics[scale=0.8]{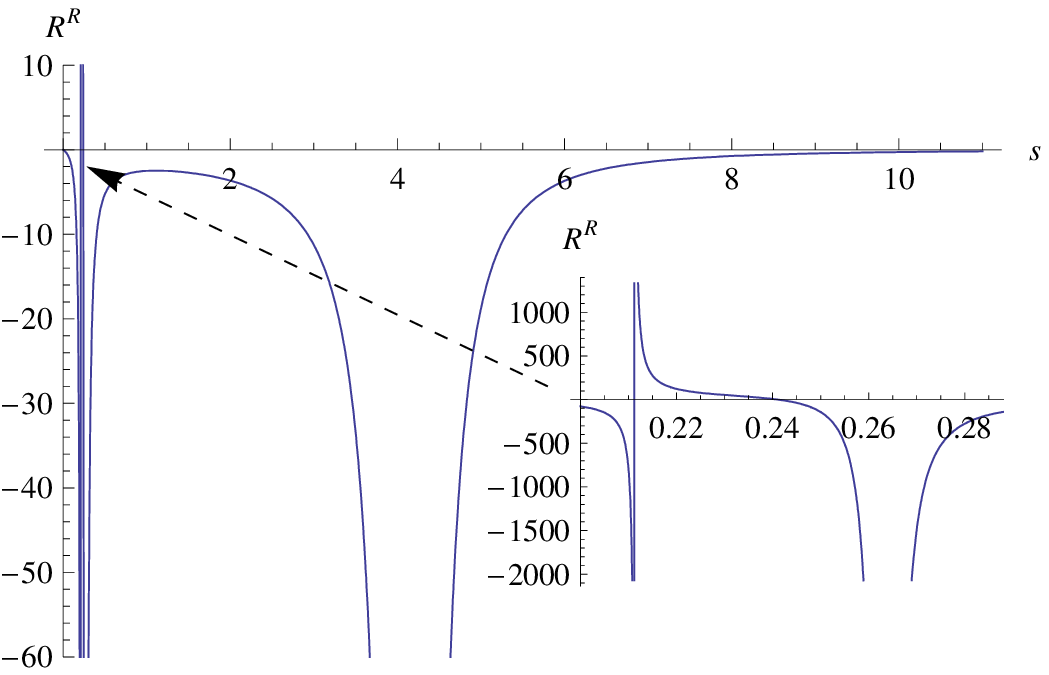}}}
\caption{The scalar curvature of thermodynamic geometry vs entropy
density  with $k=1\;,N=10,$ $q=0.4$ and $\ell_p=1\;.$ Both scalar
curvatures diverge at $s_1\approx0.264$ and $s_2\approx4.160\;.$
Note that the third singularity at $s\approx0.211$ for the scalar
curvature of Ruppeiner metric corresponds to the extremal black hole
with a vanishing Hawking temperature.}\label{curvature-WR}
\end{figure*}
\begin{figure*}[htbp]
\centering{\subfigure[$\;q=0.4<q_{crit}$]{\label{a1}
\includegraphics[scale=0.8]{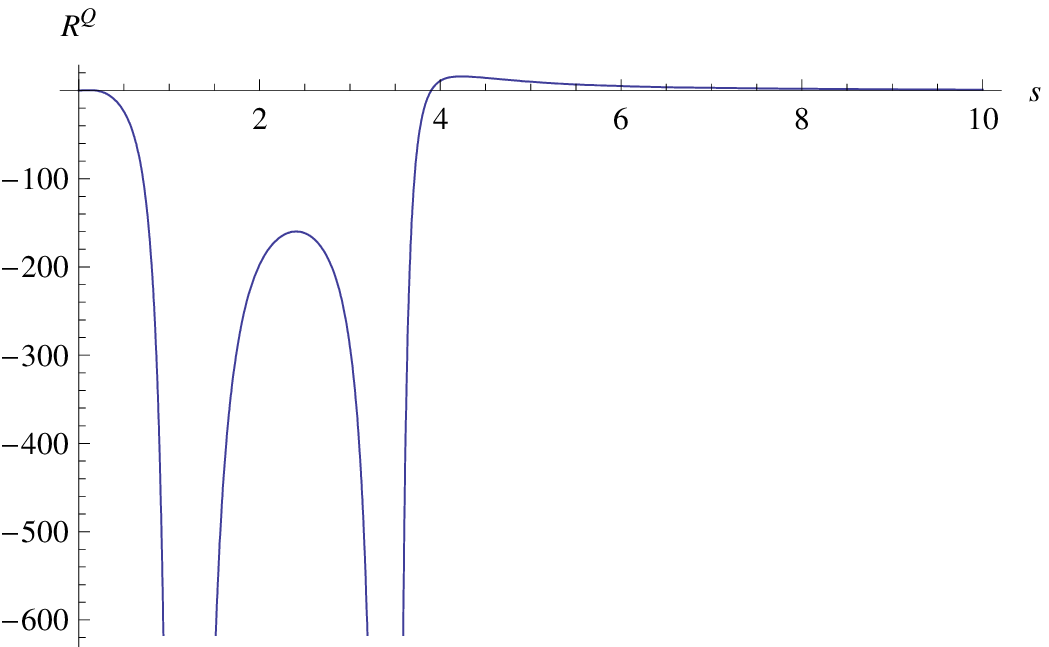}}\subfigure[$\;q=q_{crit}$]{\label{b1}\includegraphics[scale=0.8]{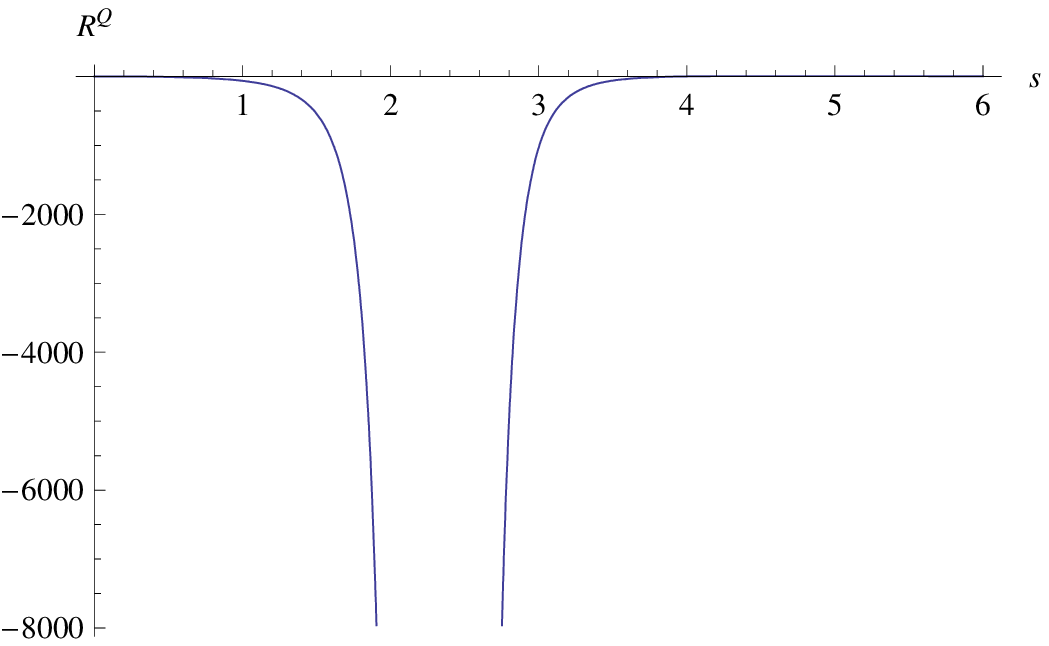}}
\subfigure[$\;q=1>q_{crit}$]{\label{b1}\includegraphics[scale=0.8]{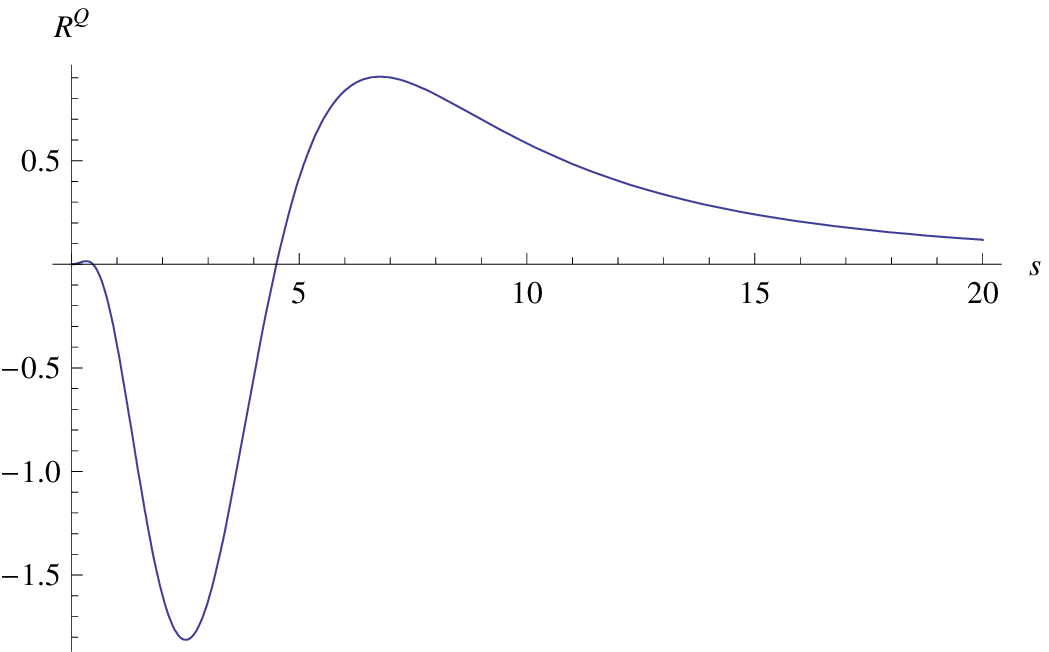}}}
\caption{The scalar curvature of Quevedo metric vs entropy density
with $k=1\;,N=10$ and $\ell_p=1\;$ for various $q\;.$  The scalar
curvature diverges at (a) $s_3\approx1.167$ and $s_4\approx3.443\;$
for $q=0.4\;,$ (b) $s_5\approx2.339\;$ for $q=q_{crit}\;,$ (c) no
divergence for $q=1>q_{crit}\;.$}\label{curvature-Q}
\end{figure*}

With the requirement for a positive Hawking
temperature~(\ref{T-uneq}),  it is easy to see that the auxiliary function $E_1(s,q)$ is always
positive. Thus the singularities of $R^R$ are just determined by the function
 $B_1(s,q)\;.$ Therefore, from Eq.~(\ref{RW}) and
Eq.~(\ref{RR}), we can conclude that both the scalar curvatures  of
the Weinhold metric and Ruppeiner metric possess the same
singularities. These singularities just coincide with the divergence
of the specific heat $C_{\Phi,N^2}$ for fixed electric potential and
numbers of colors (comparing Fig.~(\ref{cphns}) with
Fig.~(\ref{curvature-WR})). So both the Weinhold metric and
Ruppeiner metric can reveal the phase transition of the RN-AdS black
hole in the fixed $\Phi$  ensemble.

The  Quevedo metric for the RN-AdS black hole reads
\begin{equation}
g^Q=(sT+q\Phi)\Bigg(\begin{matrix}-\rho_{ss} &
0\\0&\rho_{qq}\end{matrix}\Bigg)\;.
\end{equation}
Calculating its scalar curvature gives
\begin{equation}
R^Q=\frac{A_2(s,q)}{B_2(s,q)E_2(s,q)},
\end{equation}
where $A_2(s,q)$ is a complicated regular function which we do not present here. The other  two auxiliary functions are given by
\begin{eqnarray}
B_2(s,q)&=&\big[3D\pi^2 s^{4/3} \left(2 D s^{2/3}-N^{5/6}\alpha\right)+40 N q^2 D_1 G_{10}\big]^2, \nonumber \\
 E_2(s,q)&=& \left(6 D^2 \pi^2 s^2+3 D N^{5/6} \pi^2 s^{4/3} \alpha +16 N q^2 D_1 G_{10}\right)^3. \nonumber
\end{eqnarray}
It is easy to see that $E_2(s,q)$ is a positive function and the
divergence of the scalar curvature for the Quevedo metric just
corresponds to the singular points of the specific heat $C_{q,N^2}\;$ by comparing the forms of $E_2$ and $C_{q,N^2}$
(see Fig.~(\ref{cqns}) and Fig.~(\ref{curvature-Q}) ). This means that
the Quevedo metric can reveal the phase transition of the RN-AdS
black hole  in the fixed  $q$  ensemble.

\section{ Phase transition and  thermodynamic geometry of the RN-AdS black hole in the
fixed $q$ case } \label{IV}

\begin{figure*}[htbp]
\centering{\subfigure[$\;N=5$]{\label{a1}
\includegraphics[scale=0.8]{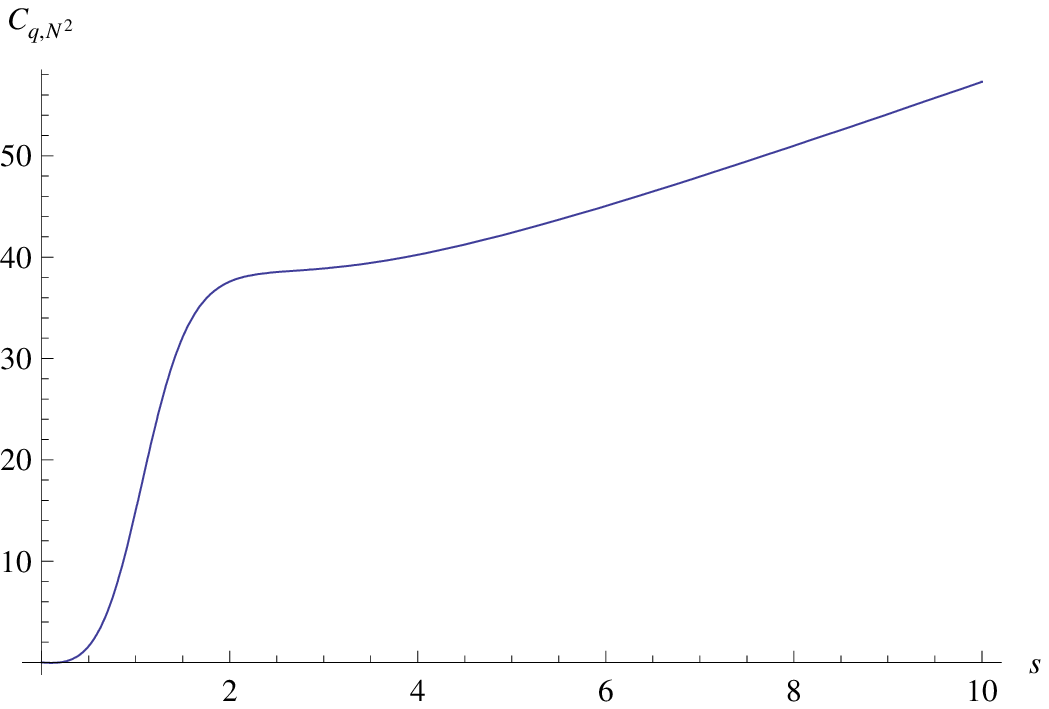}}\subfigure[$\;N=10$]{\label{b1}\includegraphics[scale=0.8]{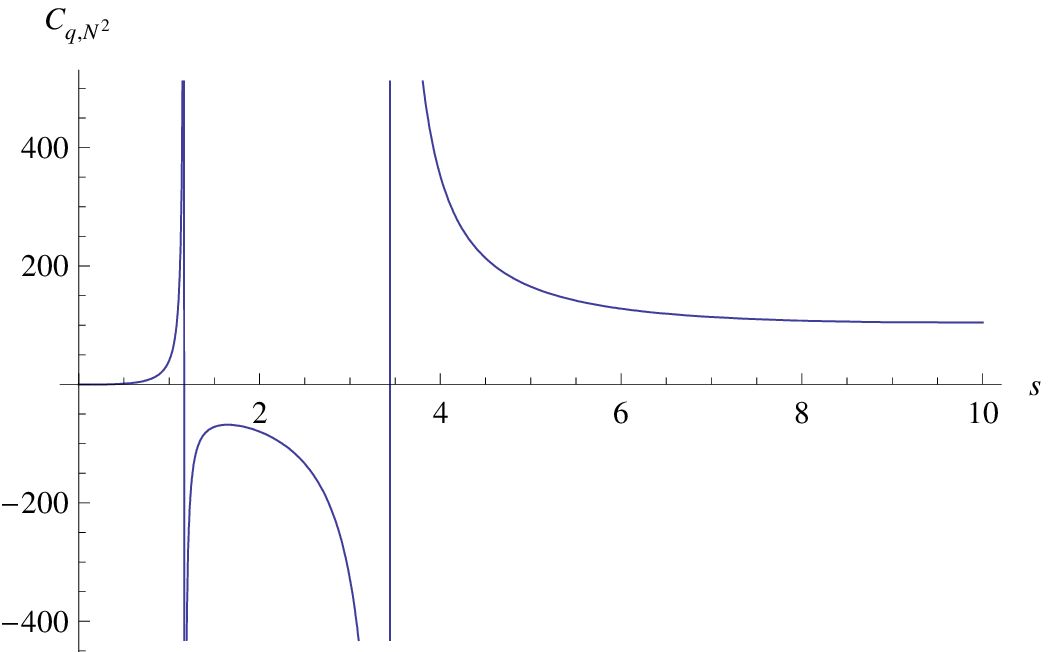}}}
 \caption{The special heat $C_{N^2,q}$  as a function of $s$  for the different
 numbers of colors with  parameters  $k=1\;,q=0.4\;$
and $\ell_p=1\;.$ Not that in this case, the critical value of the
number of colors is
$N_{crit}=\frac{12}{\alpha^2}\big(\frac{75D^2q^4D_1^2G_{10}^2}{\pi^4}\big)^{1/3}\approx7.805\;,$
so the specific heat has no divergent point in (a) for the case of
$N=5$, and   two singularities at $s_3\approx1.167$ and
$s_4\approx3.443\;$  in (b) for the case of $N=10$. }\label{cqns-2}
\end{figure*}

\begin{figure*}[htbp]
\centering{\subfigure[$\;N=10$]{\label{a1}
\includegraphics[scale=0.8]{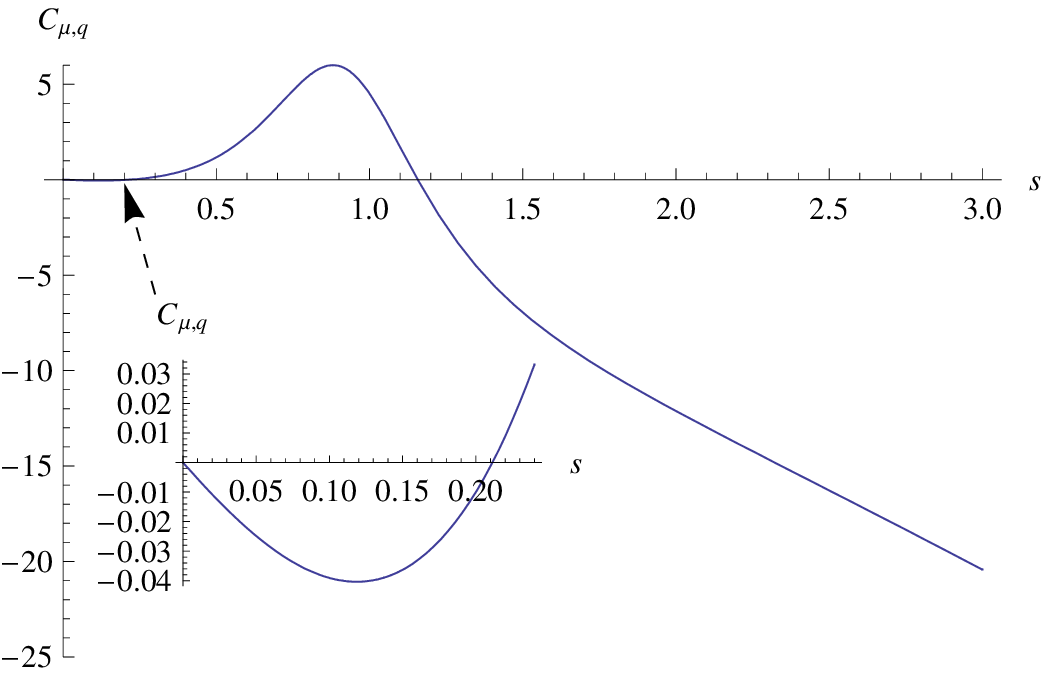}}\subfigure[$\;N=100$]{\label{b1}\includegraphics[scale=0.8]{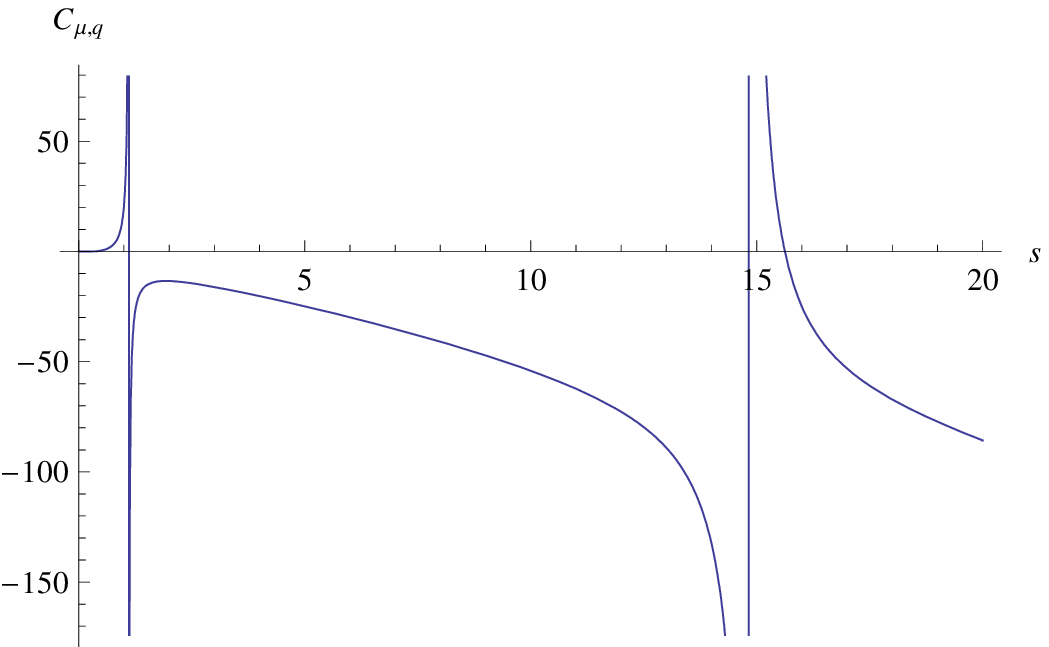}}}
 \caption{The specific heat $C_{\mu,q}$  vs  entropy  density $s$  with parameters  $k=1\;,q=0.4\;$ and $\ell_p=1\;.$  (a):  The special heat has no divergence in the case of $N=10\;$
but with  zero point at $s_6\approx1.158\;$ (note that the other
 zero point  at $s\approx0.211$
corresponds to a vanishing Hawking temperature). (b):  The specific
heat has two singularities  at $s_7\approx1.106$ and
$s_8\approx14.825\;$ and a physical  zero point  at
$s\approx15.624\;$ in the case of $N=100\;.$}\label{cmuqs}
\end{figure*}

 Now we turn to the fixed $q$ case. This means that the charge density $q$ is treated as a fixed external parameter, not a thermodynamic variable.
  Then the corresponding
specific  heats  are  $C_{N^2,q}$ and
$C_{\mu,q}\;.$ The exact expression  of $C_{N^2,q}$ is  given
in Eq.~(\ref{c-nq}) in the previous section, from which we can see there is a critical value $N_{crit}$,
for the number of colors. When
$N<N_{crit}= \frac{12}{\alpha^2}\big(\frac{75D^2q^4D_1^2G_{10}^2}{\pi^4}\big)^{1/3}\;,$
there does not exist any divergence for the specific  heat
$C_{N^2,q}$. On the other hand, when $N >N_{crit}$,  there exist some divergences for this specific heat, which indicates  some phase transition  in
 canonical ensemble with a fixed $q$. We plot the behavior of $C_{N^2,q}$ as
a function of $s$ in Fig.~(\ref{cqns-2}) with fixed $q=0.4$ for
two values of the number of colors to show the feature.
On the other hand, the  specific heat  $C_{\mu,q}\;$ with a fixed chemical potential can be calculated as
 \begin{equation} \label{c-muq}
C_{\mu,q}=T\bigg(\frac{\partial{s}}{\partial{T}}\bigg)_{\mu,q}=\frac{C_1(s,N)C_2(s,N)}{C_3(s,N)}\;,
 \end{equation}
where the auxiliary functions read
\begin{eqnarray}
C_1(s,N)&=&3D\pi^2 s^{4/3}\big(64 D s^{2/3}-11 N^{5/6}
\alpha \big)-160 N q^2 D_1 G_{10}\;,
 \nonumber \\
C_2(s,N)&=&-\frac{s}{9}\big[3 D \pi ^2 s^{4/3} \big( 2 D
s^{2/3}+N^{5/6} \alpha \big)-8 N q^2 D_1 G_{10}\big]\;,
\nonumber\\
C_3(s,N)&=&3 D^2 N^{5/6}\pi^4 s^{8/3}\alpha\big(6 D s^{2/3} -
    N^{5/6}\alpha \big) + 256 N^2 q^4D_1^2 G_{10}^2 \nonumber\\
    &&-24 D N\pi^2 q^2 s^{4/3}D_1 G_{10}\big(8 D s^{2/3} -
    N^{5/6}\alpha \big)\;.\nonumber
\end{eqnarray}
Having considered the positiveness  of the Hawking
temperature~(\ref{T-uneq}), we find that  $C_2(s,N)$ is always
negative. The zero point of the specific heat is only  relevant to
the function $C_1(s,N)\;,$ while the  singularities of the specific
heat are determined by the function $C_3(s,N)\;.$ We show the
behavior of $C_{\mu,q}$ as a function of $s$ in Fig.~(\ref{cmuqs}).
An interesting observation from the figure is that  $C_{\mu,q}$  has
a chance  to be zero even for a non-extremal black hole. But the
implication of this is not yet clear.

 In the fixed $q$ case, the Weinhold metric is two-dimensional and can be expressed as
\begin{equation}\label{gW}
g^W=\Bigg(\begin{matrix}\rho_{ss} & \rho_{sN^2}\\
{\rho_{N^2s}}&{\rho}_{N^2N^2}\end{matrix}\Bigg)\;.
\end{equation}
 The corresponding  scalar curvature reads
 \begin{equation}
R^W=A_3(s,N)/B_3(s,N)\;,
 \end{equation}
 where $A_3(s,N)$ is a complicated regular function which we do not present here, while
 \begin{eqnarray}
 B_3(s,N)=N^{-3/2}[C_3(s,N)]^2\;.
 \end{eqnarray}
And the  scalar curvature of Ruppeiner metric can also
 be obtained as
 \begin{equation}
R^R=\frac{A_4(s,N)}{B_4(s,N)},\;
 \end{equation}
with
 \begin{equation}
 B_4(s,N)= 27 C_2(s,N)[ C_3(s,N)]^2, \;
\end{equation}
and again $A_4(s, N)$ is a regular function of $s$ and $N$.
\begin{figure*}[htbp]
\centering{\subfigure[Weinhold metric]{\label{a1}
\includegraphics[scale=0.8]{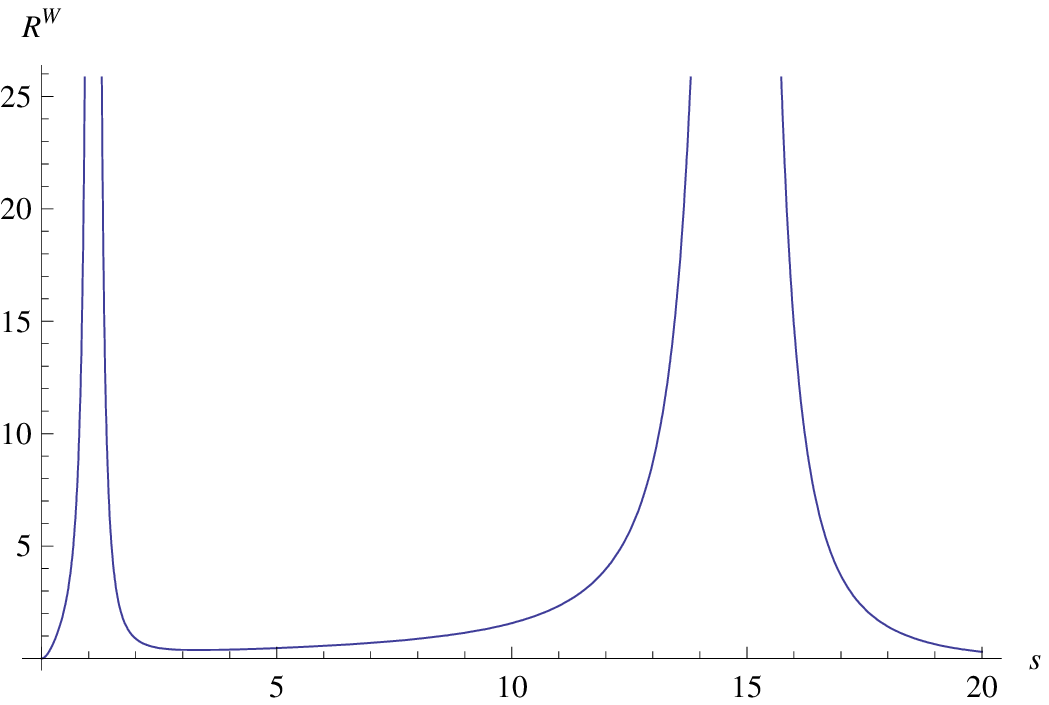}}\subfigure[Ruppeiner metric]{\label{b1}\includegraphics[scale=0.8]{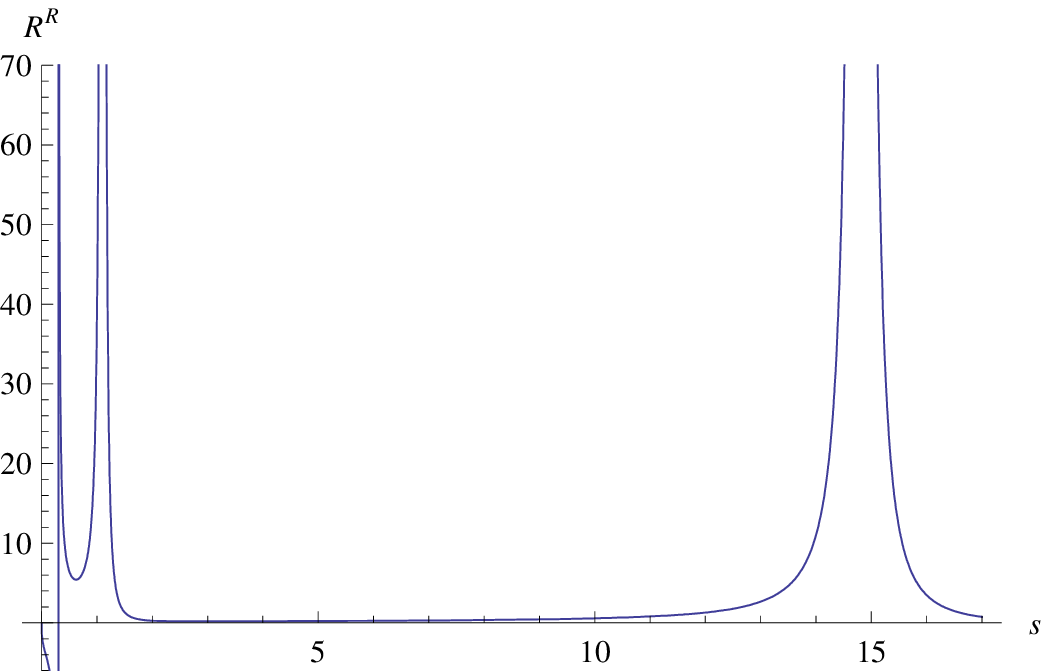}}}
 \caption{The scalar curvature of thermodynamical geometry  vs  entropy density for the (a) Weinhold
 metric and (b) Ruppeiner metric  with parameters  $k=1\;,q=0.4\;,N=100$
and $\ell_p=1\;.$ Both the scalar curvatures diverge at
$s_7\approx1.106$ and $s_8\approx14.825\;,$    the same points of
divergence  of the specific heat $C_{\mu,q}$ (see plot (b)
in Fig.~(\ref{cmuqs})). Note that the other singularity in plot
(b)  corresponds to the extremal black hole with a vanishing Hawking temperature.
}\label{curvature2-WR}
\end{figure*}
 In Fig.~(\ref{curvature2-WR}), we plot the scalar
curvature for the Weinhold metric  and Ruppeiner metric  as a function of entropy density with parameters
$q=0.4\;,N=100\;$ and $\ell_p=1\;.$ It is easy to see that  both the
Weinhold metric and Ruppeiner metric can give correct information of phase transition for the RN-AdS black hole in the  fixed $q$  case
  in the fixed chemical potential $\mu$  ensemble. Namely the singularity behavior of the thermodynamical curvature for both the Weinhold
  metric and Ruppeiner metric is consistent with  that of $C_{\mu,q}$.  Note that there is an additional singularity in plot (b) in Fig.~(\ref{curvature2-WR}), which
  corresponds to the extremal black hole with a vanishing Hawking temperature.

\begin{figure*}[htbp]
\centering{\subfigure[]{\label{a1}
\includegraphics[scale=0.8]{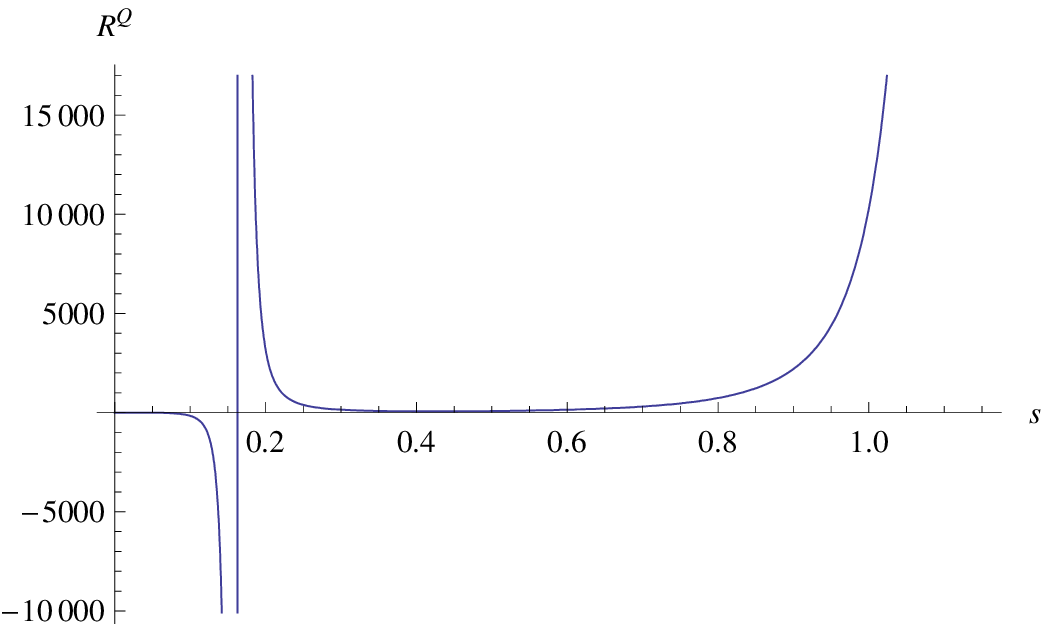}}\subfigure[]{\label{b1}\includegraphics[scale=0.8]{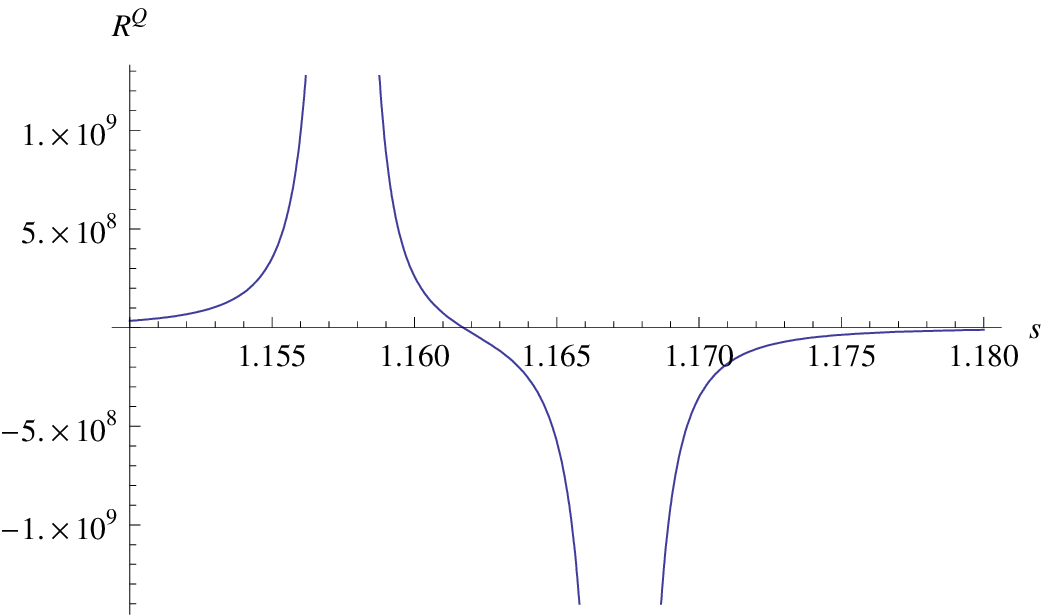}}
\subfigure[]{\label{b1}\includegraphics[scale=0.8]{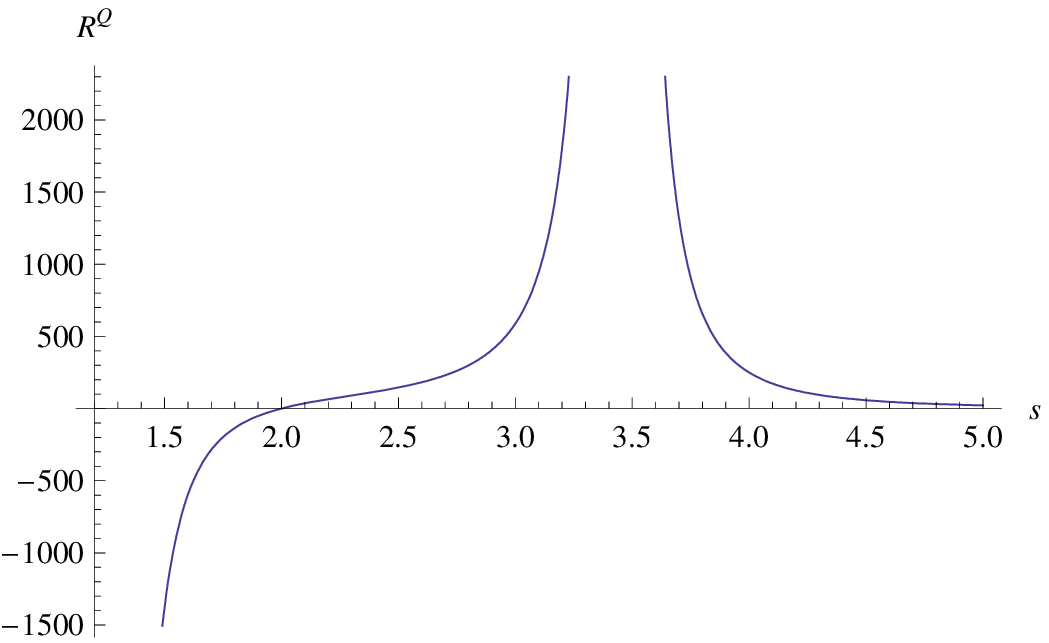}}}
 \caption{The  scalar curvature  as a function
of entropy density in the Quevedo metric with parameters
$k=1\;,q=0.4\;,N=10$ and $\ell_p=1$. There exist three physical
divergences at $s_6 \approx 1.158$, $s_3 \approx 1.167$ in plot (b)
and $s_4 \approx 3.443$ in plot (c), respectively. And the remaining
one at $s=0.162$ in plot (a) corresponds a negative Hawking
temperature, thus it is not physical relevant. }\label{curvature2-Q}
\end{figure*}

In the case with a fixed $q$,  the Quevedo metric reads
\begin{equation}
g^Q=(sT+N^2\mu)\Bigg(\begin{matrix}-\rho_{ss} &
0\\0&\rho_{N^2N^2}\end{matrix}\Bigg)\;.
\end{equation}
The corresponding scalar curvature can be expressed as
\begin{equation}\label{RR-2}
R^R=\frac{A_5(s,N)}{[C_1(s,N)]^2E_5(s,N)B_5(s,N)}\;,
\end{equation}
where  $A_5(s,N)$ is a regular function, while the other two
auxiliary functions  are given by
\begin{equation}
E_5(s,N)=\big[3D\pi^2s^{4/3}\big(4 D s^{2/3}+3 N^{5/6}\alpha
\big)-16 Nq^2D_1G_{10}\big]^3\;,
\end{equation}
and
\begin{equation}
B_5(s,N)=\big[ 3D\pi^2 s^{4/3} \big(2D s^{2/3}-N^{5/6}\alpha
\big)+40 N q^2 D_1 G_{10}\big]^2\;.
\end{equation}
With a positive  Hawking temperature, the
auxiliary function $E_5(s,N)$ is always positive.  Then  the
divergence of the scalar curvature is related to the behaviors of $B_5(s,N)$ and
$C_1(s,N)$.  Comparing Eq.~(\ref{RR-2}) with Eq.~(\ref{c-nq})
and Eq.~(\ref{c-muq}), we can  conclude that the singularity behavior of the scalar
curvature is the same as that of the specific heat of $C_{N^2,q}$, which means that
 the scalar  curvature can reveal the information of phase transition of the black hole
 in the fixed $N$ ensemble.  In addition, let us note that there exists an additional  singularity
 arising from the auxiliary function $C_1(s,N)$(e.g., $s_6 \approx 1.158$ in Fig.~(\ref{curvature2-Q})), which corresponds to the zero point of the specific heat $C_{\mu,q}$.
We can see these behaviors by  comparing Fig.~(\ref{curvature2-Q})
with plot (b) in Fig.~(\ref{cqns-2}) and  plot (a) in
Fig.~(\ref{cmuqs}). These results are in agreement  with the recent
study in \cite{Hendi,Suresh} that  the divergences of the scalar
curvature for the Quevedo metric correspond to the singularities  or
zero for some specific heats.

\section{Conclusions}
\label{V}

In this paper, we have studied the thermodynamics of a RN-AdS black hole
 in the extended phase space where
the cosmological constant is related to the number of colors in the
dual Supersymmetric Yang-Mills theory. Note that  this extended phase space is
different from the one where the cosmological constant acts as the pressure and its conjugate as the thermodynamical
volume.  In the former case, the mass of the black hole is viewed as the internal energy of the thermodynamical system, while
in the latter case, the mass of the black hole acts as the enthalpy~\cite{Dolan1,Dolan2}.  We calculated and discussed
the chemical potential associated with the number of colors, and
found that the contribution of the charge of the black hole to chemical
potential is always positive. The chemical potential has a chance to
be positive, and the existence of charge make the chemical potential
become positive more easily. In the figure of the chemical potential as a function of temperature, we see
that there exists a region where the chemical potential is a multi-valued function. This region just corresponds
to the unstable branch of the black hole with a negative heat capacity.  The chemical potential is found to be negative
for large black holes, while it is positive for small ones.  This behavior is  qualitatively the same as that in the case of the Schwarzschild-AdS black holes~\cite{ZCY}.

 The corresponding specific heats have been calculated  respectively in
 the fixed $N^2$ case and the fixed $q$ case.  It has been found that the specific heat in the fixed number of colors and fixed
 charge density $C_{N^2,q}$ has a certain critical  point, if the charge is larger
 than its critical value or the number of colors is smaller than its critical value,  then $C_{N^2,q}$  has no
 divergence. The specific heat for a fixed electric potential and fixed number of colors
 $C_{\Phi,N^2}$ can not be zero for non-extremal black holes, while
 the specific heat for the fixed chemical potential and fixed charge
 $C_{\mu,q}$ has a chance to be zero even for a non-extremal black hole.

In the extended phase space, we have studied the thermodynamical
geometry associated with the RN-AdS black hole. By calculating
scalar curvatures of the Weinhold metric, Ruppeiner metric and
Quevedo metric in the fixed $N^2$ case, we found that in the
Weinhold metric and Ruppeiner metric both the scalar curvatures
diverge at the same divergent points of $C_{\Phi,N^2}$, while in the
Quevedo metric, the scalar curvature diverges at the singularity  of
$C_{q,N^2}$.  In the fixed $q$ case, both the scalar curvatures in
the Weinhold metric and Ruppeiner metric diverge at the  divergent
points of $C_{\mu,q}\;,$ while the scalar curvature in the Quevedo
metric diverges at  the point of $C_{\mu,q}=0$, besides  the
divergent points of $C_{N^2,q}.$ These results indicate that the
divergence of thermodynamical curvature indeed is related to some
divergence of specific heats, but the divergence of thermodynamical
curvature might  be also relavent to the vanishing points of the
specific heat. These studies are helpful to further understand the
relation between phase transition and divergence of thermodynamical
curvature. To further study this relation, it should be of great
interest to discuss thermodynamics and thermodynamical curvature for
other black holes in AdS space in the extended phase space. In
particular, it is called for to further understand the potential
role associated with the cosmological constant in the dual field
theory according to the AdS/CFT correspondence.

\begin{acknowledgments}
R.G.C thanks Hunan Normal University for the warm hospitality
extended to him during his visit.
 This work was supported in part by the National Natural Science
Foundation of China ( N0.11005038,No.10975168,
No.11035008,No.11375092 and No.11435006), the Hunan Provincial
Natural Science Foundation of China under Grant No. 11JJ7001 and the
Program Excellent Talent Hunan Normal University No. ET13102.
\end{acknowledgments}

\end{document}